\documentclass[aps,pra,twocolumn,superscriptaddress,floatfix,
nofootinbib,showpacs,longbibliography]{revtex4-1}
\usepackage[utf8]{inputenc}  
\usepackage[T1]{fontenc}     
\usepackage[british]{babel}  
\usepackage[sc,osf]{mathpazo}
\usepackage[scaled=0.86]{berasans}  
\usepackage[colorlinks=true, citecolor=blue, urlcolor=blue]{hyperref}  
\usepackage{graphicx} 
\usepackage[babel]{microtype}  
\usepackage{amsmath,amssymb,amsthm,bm,amsfonts,mathrsfs,bbm} 

\usepackage{xspace}  
\usepackage{pgf,tikz}
\usepackage{xcolor}
\usepackage{multirow}
\usepackage{array}
\usepackage{bigstrut}
\usepackage{braket}
\usepackage{color}
\usepackage{natbib}
\usepackage{multirow}
\usepackage{mathtools}
\usepackage{float}
\usepackage{xcolor,colortbl}
\usepackage{color}
\usepackage[justification=justified, format=plain]{subcaption}
\usepackage[justification=raggedright]{caption}

\newcommand{\be}{\begin{equation}}
\newcommand{\ee}{\end{equation}}
\newcommand{\ba}{\begin{eqnarray}}
\newcommand{\ea}{\end{eqnarray}}

\def\>{\rangle}
\def\<{\langle}

\begin{document}
	
\title{Activating information backflow with the assistance of quantum SWITCH}	

\author{Ananda G. Maity}
\email{anandamaity289@gmail.com}
\affiliation{Networked Quantum Devices Unit, Okinawa Institute of Science and Technology Graduate University, Onna-son, Okinawa 904-0495, Japan}

\author{Samyadeb Bhattacharya}
\email{samyadeb.b@iiit.ac.in}
\affiliation{Center for Security Theory and Algorithmic Research,\\
Center for Quantum Science and Technology,\\
International Institute of Information Technology, Gachibowli, Hyderabad 500 032, India}
	
\begin{abstract}
There are certain dynamics while being non-Markovian, do never exhibit information backflow. We show that if two such dynamical maps are considered in a scenario where the order of application of these two dynamical maps are not definite, the effective channel can manifest information backflow. In particular, we use quantum SWITCH to activate such a channel. In contrast, activation of those channels are not possible even if one uses many copies of such channels in series or in parallel action. We then investigate the dynamics behind the quantum SWITCH experiment and find out that after the action of quantum SWITCH both the CP (Complete Positive)- divisiblity and P (Positive)- divisibility of the channel breaks down, along with the activation of information backflow. Our study elucidate the advantage of quantum SWITCH by investigating its dynamical behaviour.
\end{abstract}

\maketitle

\section{Introduction}
The theory of open quantum system deals with the system interacting with noisy environment and hence essentially serves as an effective tool for calculating the dynamics of a many body system \cite{breuer,alicki,lindblad,gorini,rivas1,breuerN,alonso}. This is applicable widely since no physical system is truly isolated. For a general Markovian evolution, the system interacts weakly with large and stationary environment and hence the dynamics are considered to be memoryless leading to one way information flow from the system to the environment. Therefore, the quantumness of a system subjected to Markovian dynamics eradicates gradually with time \cite{RHPreview,BLPreview,Vegareview}. However, in the realistic scenario, the system-environment coupling may not be sufficiently weak and the environment can be finite as well as non-stationary, which leads to the signature of non-Markovian backflow of information \cite{RHP,blp1,bellomo,arend,ban1,ban2,Bhattacharya17,samya2,Bhattacharya20,Maity20,BBhattacharya21,D'Arrigo14,Hsieh19,Frigerio21}.

In recent times, non-Markovian information backflow has been shown to be a useful resource for several information processing tasks like perfect teleportation with mixed states \cite{task1}, distribution of entanglement efficiently \cite{task3}, improving the capacity for long quantum channels \cite{task2}, extracting work from an Otto cycle \cite{task4}, controlling quantum system \cite{task5} and so on.

In view of this enormous applications of non-Markovianity, it is therefore important to analyse the nature of system-environment dynamics that triggers non-Markovian traits. Usually, the signature of quantum non-Markovian dynamics are characterized and exposed by the properties of {\it indivisibility} \cite{RHP,Bhattacharya20}. A dynamical map $\Lambda (t,t_0)$ is said to be CP-divisible if it can be represented as $\Lambda (t, t_0) = \Lambda (t,t_s) \circ \Lambda (t_s,t_0)$, $\forall t_s$ with $t_0 \leq t_s \leq t$ and both $\Lambda (t,t_s)$ and $\Lambda (t_s, t_0)$ are completely positive. However, if the dynamical map can not be written in the above form for at least some $t_s$, then the dynamics will be indivisible. For any indivisible dynamics, although the overall dynamics is CP \cite{choi,jamil}, it may not be CP for certain intermediate time steps. Here it is important to mention that though the definition of quantum non-Markovianity has been recently claimed to be more general \citep{modi1,modi2,modi3,modi4}, we are considering the divisibility based definition for the purpose of this particular work.

To this end, it is important to discuss two main approach towards quantum non-Markovianity; i.e. CP divisibility \citep{RHP} and information backflow \citep{blp1}. CP divisibility, as defined earlier, is the property of a channel which allows to realize it as concatenations of infinite CP maps; i.e. the dynamics can be divided into infinite such dynamical maps, whereas information backflow is a property of CP indivisible maps which allows anomalous increment of distance like measures that are monotones under divisible CP dynamics. More explicitly, the phenomena of information backflow will take place if and only if there exists atleast one time instant $s$ and one pair of initial states ($\rho^a (s)$ and  $\rho^b (s)$) for which the distance between $\rho^a (s)$ and  $\rho^b (s)$ grows. It has been shown earlier \citep{stromer} that information backflow is a sufficient condition for CP indivisibility of a quantum channel, but not necessary. Hence, although for most of the evolutions, information backflow is the signature of CP-indivisibility, not all indivisible (and hence non-Markovian) dynamics shows information backflow. Therefore, for those particular dynamics, most of the proposed measure of non-Markovianity are not sensitive and hence the signature of non-Markovianity for those dynamics may be undetected with that particular measures. Few of those dynamics, where non-Markovianity is hidden with respect to the measures based on information backflow can be found in the well known examples of {\it eternally non-Markovian} quantum channels \cite{Hall14,Samya_eternal,eternal1,eternal2,sabrina}. It is worth highlighting that this form of non-Markovianity, even though concealed concerning information backflow, fundamentally deviates from the concept of `hidden non-Markovianity' introduced by Burgarth {\it et. al.} \cite{Burgarth21}. In their work, it has been demonstrated that non-Markovian open quantum systems can exhibit precisely Markovian dynamics for an arbitrarily long time. Consequently, the non-Markovian nature of these systems remains entirely `hidden', implying that it cannot be experimentally detectable by observing the dynamics within a finite, albeit large, time frame.

Recently, there has been substantial interest to study whether the causal order of two events can be made indefinite and if so, upto what extent such indefiniteness can be exploited as a resource in several information processing tasks. Although the concept was initially proposed by Lucian Hardy \cite{Hardy05}, however, its information theoretic application was first pointed out by Chribella {\it et. al} \cite{Chiribella13}. Indeed, they introduced the concept of quantum SWITCH where an additional ancillary system is used to control the order of the action of two quantum operations $\mathcal{E}_1$ and $\mathcal{E}_2$ on some quantum system say $\rho$. For example when the control bit is initialised in $\ket{0}$, first operation $\mathcal{E}_1$ will be applied and then $\mathcal{E}_2$ on $\rho$. On the other hand when the control bit is initialised in $\ket{1}$, the action of these two quantum operations will be in reverse order i.e, first operation $\mathcal{E}_2$ and then operation $\mathcal{E}_1$ will be carried out on the state $\rho$. Now, one can in principle initialise this control qubit in superposition basis and thus the order of the action of two quantum operations can be made indefinite. A more general and stronger notion of such causal indefiniteness was later proposed by Oreshkov {\it et. al.} \cite{Oreshkov12} via {\it process matrix} formalism. Thereafter, indefinite causal order has been shown to be extensively useful in several information processing and communication protocols such as testing the properties of quantum channel \cite{Chiribella12}, winning non-local games \cite{Oreshkov12}, enhancing the precision of quantum  metrology \cite{Zhao20}, improving quantum communication \cite{Ebler18,Salek18,Chiribella21}, minimizing quantum communication complexity \cite{Guerin16}, achieving quantum computational \cite{Araujo14} and thermodynamic \cite{Tamal20,Vedral20} advantages and so on \cite{Mukhopadhyay19,Banik21,Maity23}. Very recently, indefinite causal order and its advantage has also been realised experimentally \cite{Procopio15,Rubino17,Goswami18(1)}.

Motivated by the above fact, here we ask the question whether we can utilize quantum SWITCH to activate the information backflow of a channel which does not show any backflow of information initially. In order to do so, we take two copies of such channels containing eternal non-Markovianity that do not manifest any information backflow and then use them in superposition of two different orders using an extra ancillary qubit system, controlling the order of actions of the channels. After measuring the ancillary system in coherent basis, we show that the effective channel can in general be activated for information backflow. Therefore, after the action of quantum SWITCH, those non-Markovian dynamics can also be detected through the usual measures of information backflow. 

\section{A brief review on quantum-SWITCH}\label{s2}
In a fixed causal order, one can use the channels in either series or parallel combination. Given two channels $\mathcal{N}_1$ and $\mathcal{N}_2$, the parallel combination of the channels can be represented by the tensor product structure as $\mathcal{N}_1\otimes \mathcal{N}_2$. On the other hand, in case of series combination, two channels can be composed in two distinct ways: either $\mathcal{N}_2 \circ \mathcal{N}_1$ or $\mathcal{N}_1 \circ \mathcal{N}_2$. In the former, the channel $\mathcal{N}_1$ acts first, followed by $\mathcal{N}_2$, while in the latter, the order of channel action is reversed.
 
 One can in principle control the order of the action of two channels by using an additional ancillary system called the {\it control bit}. In particular, one can use a control bit $\ket{0}$ when the channel $\mathcal{N}_1$ is applied before the channel $\mathcal{N}_2$ (i.e, $\mathcal{N}_2 \circ \mathcal{N}_1$) and use $\ket{1}$ when the channel $\mathcal{N}_1$ is applied after the channel $\mathcal{N}_2$ (i.e, $\mathcal{N}_1 \circ \mathcal{N}_2$). If the Kraus operator corresponding to channel $\mathcal{N}_1$ and $\mathcal{N}_2$ are $\lbrace K_i^{(1)}\rbrace$ and $\lbrace K_j^{(2)} \rbrace$ respectively, then the Kraus operators of the overall quantum channel resulting from the switching of two channels $\mathcal{N}_1$ and $\mathcal{N}_2$ will be as follows,
 \begin{equation}
     W_{ij} = K_j^{(2)} \circ K_i^{(1)} \otimes \ket{0}_c\bra{0} + K_i^{(1)} \circ K_j^{(2)} \otimes \ket{1}_c\bra{1}.
 \end{equation}
 This joint Kraus acts over the target state $\rho$ and the control bit $\omega_c$. Therefore, the overall evolution of the joint system (target and control) will be
 \begin{equation}
     S(\mathcal{N}_1,\mathcal{N}_2)(\rho \otimes \omega_c) = \sum_{i,j} W_{ij} (\rho \otimes \omega_c) W_{ij}^{\dagger}.
 \end{equation}
 Finally, the control qubit is measured on the coherent basis $(\lbrace \ket{+}\bra{+}, \ket{-}\bra{-}\rbrace)$ and the effective subsystem (target state) is obtained corresponding to each outcome as
  \begin{equation}
     {}_c\bra{\pm}S(\mathcal{N}_1,\mathcal{N}_2)(\rho \otimes \omega_c) \ket{\pm}_c.
 \end{equation}
\begin{figure}[htb]
\centering
\includegraphics[height=5cm,width=8.5cm]{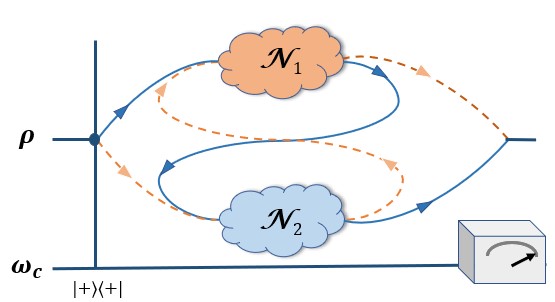}
\caption{The solid blue line represents that the quantum state $\rho$ is first traversing through channel $\mathcal{N}_1$ and then through $\mathcal{N}_2$ and the dotted orange line represents that the state $\rho$ is first traversing through channel $\mathcal{N}_2$ and then through $\mathcal{N}_1$. The former take places if the control bit, $\omega_c$ is initialised in $\ket{0}_c\bra{0}$ state and the later take places if the control bit is initialised in $\ket{1}_c\bra{1}$ state. Now if $\omega_c = \ket{+}_c\bra{+}$, then the state $\rho$ will be sent to the receiver's end through the effective channel made of superposition of $\mathcal{N}_1 \circ \mathcal{N}_2$ and $\mathcal{N}_2 \circ \mathcal{N}_1$. Finally the control qubit is accessed through measurement.}
\end{figure}

\section{Eternal non-Markovianity and its activation through quantum-SWITCH}\label{s3}
Let us now consider the dynamics of Lindblad form, $\dot{\rho}(t) = \mathcal{L}_t (\rho (t))$ where $\mathcal{L}_t (X) = \sum_i \Gamma_i(t)(A_i X A_i^{\dagger} - \frac{1}{2}\lbrace A_i^{\dagger}A_i, X\rbrace)$ with $\Gamma_i(t)$ being the Lindblad coefficients and $A_i$ being the Lindblad operators. For differentiable dynamical maps \cite{Li2018}, the necessary and sufficient condition for a operation being CP-divisible is that all the Lindblad coefficients $\Gamma_i(t)$ are positive $\forall (i,t)$ \cite{gorini}. However, in realistic scenario, when the system environment coupling is not sufficiently weak, CP-divisibility may breakdown. While this breakdown of CP-divisibility in general leads to the backflow of information from the environment to the system, there are certain situation where even though CP-divisibility breaks down, information backflow may not take place. Examples of such types of evolution can be found in {\it eternally non-Markovian} quantum channels. Let us now contemplate one of the simplest scenario which captures the phenomenon of {\it eternal non-Markovianity} \cite{Hall14}. 

Consider the following qubit master equation
\begin{equation}\label{depol1}
    \frac{d}{dt}\rho (t)= \mathcal{L}(\rho (t))= \sum_{i=1}^3 \frac{\gamma_i(t)}{2}[\sigma_i \rho (t) \sigma_i - \rho (t)],
\end{equation}
where $\gamma_i (t)$ are the time-dependent Lindblad coefficients and for our specific purposes we choose $\gamma_1(t) = \gamma_2(t) = 1$ and $\gamma_3 (t)= -\tanh{t}$ \cite{Hall14}, $\sigma_i$ are the Pauli matices and the qubit is represented by
\[\rho(t) = \left(\begin{array}{cc} \rho_{11}(t)  & \rho_{12}(t) \\ \rho_{21}(t) & \rho_{22}(t) \end{array}\right).\] 
The corresponding dynamics can be represented by the following completely positive trace preserving map $\Lambda (t,0)$, such that $\rho (t) = \Lambda (t,0) \rho (0)$. More elaborately,
\begin{align}\label{5}
    \rho_{11}(t)&= \rho_{11} (0) \left( \frac{1+e^{-2\xi_1(t)}}{2}\right) + \rho_{22} (0)\left( \frac{1-e^{-2\xi_1(t)}}{2}\right), \nonumber \\
    \rho_{12}(t)&= \rho_{12}(0)e^{-\xi_2(t)} \nonumber \\
    \rho_{22}(t) &= 1-\rho_{11}(t),
\end{align}
with $\xi_1(t)=\int_0^t \gamma_1(s) ds$ and $\xi_2(t)=\int_0^t (\gamma_1(s)+ \gamma_3(s))ds$.  Since one of the Lindblad coefficient is always negative, such a map is not Markovian. Despite having a eternally negative Lindblad coefficient, the dynamics is overall a valid CP evolution, though certainly not divisible. However, the channel or dynamical map does not show information backflow and as a result, such instances of non-Markovianity cannot be detected by several well-known monotone based measures of non-Markovianity \cite{RHP,blp1}. This is because even though the dynamics is CP-divisible, it always retains its P-divisibility, i.e. all intermediate maps, while certainly not CP, remain positive. Hence, several distance measures like trace distance, retain their monotonic property under a positive trace preserving map \citep{Hall14}. 

In principle, the Kraus operators corresponding to this dynamical map can be written as follows:
\begin{align}
    K_1 &= \sqrt{A_2(t)} \left(\begin{array}{cc} 0  & 1 \\ 0 & 0 \end{array}\right), ~~~K_2 = \sqrt{A_2(t)} \left(\begin{array}{cc} 0  & 0 \\ 1 & 0 \end{array}\right), \nonumber \\
    K_3 &= \sqrt{\frac{A_1(t)+A_3(t)}{2}} \left(\begin{array}{cc} 1  & 0 \\ 0 & 1 \end{array}\right), \nonumber \\
    K_4 &= \sqrt{\frac{A_1(t)-A_3(t)}{2}} \left(\begin{array}{cc} -1  & 0 \\ 0 & 1 \end{array}\right).
\end{align}
with $A_1(t) = A_3(t)=\frac{1+e^{-2t}}{2}$ and $A_2(t) =  \frac{1-e^{-2t}}{2}$ which can be evaluated from Eq. \eqref{5}. This also implies $K_4=0$.

\begin{figure}[htb]
\centering
\includegraphics[height=5cm,width=8cm]{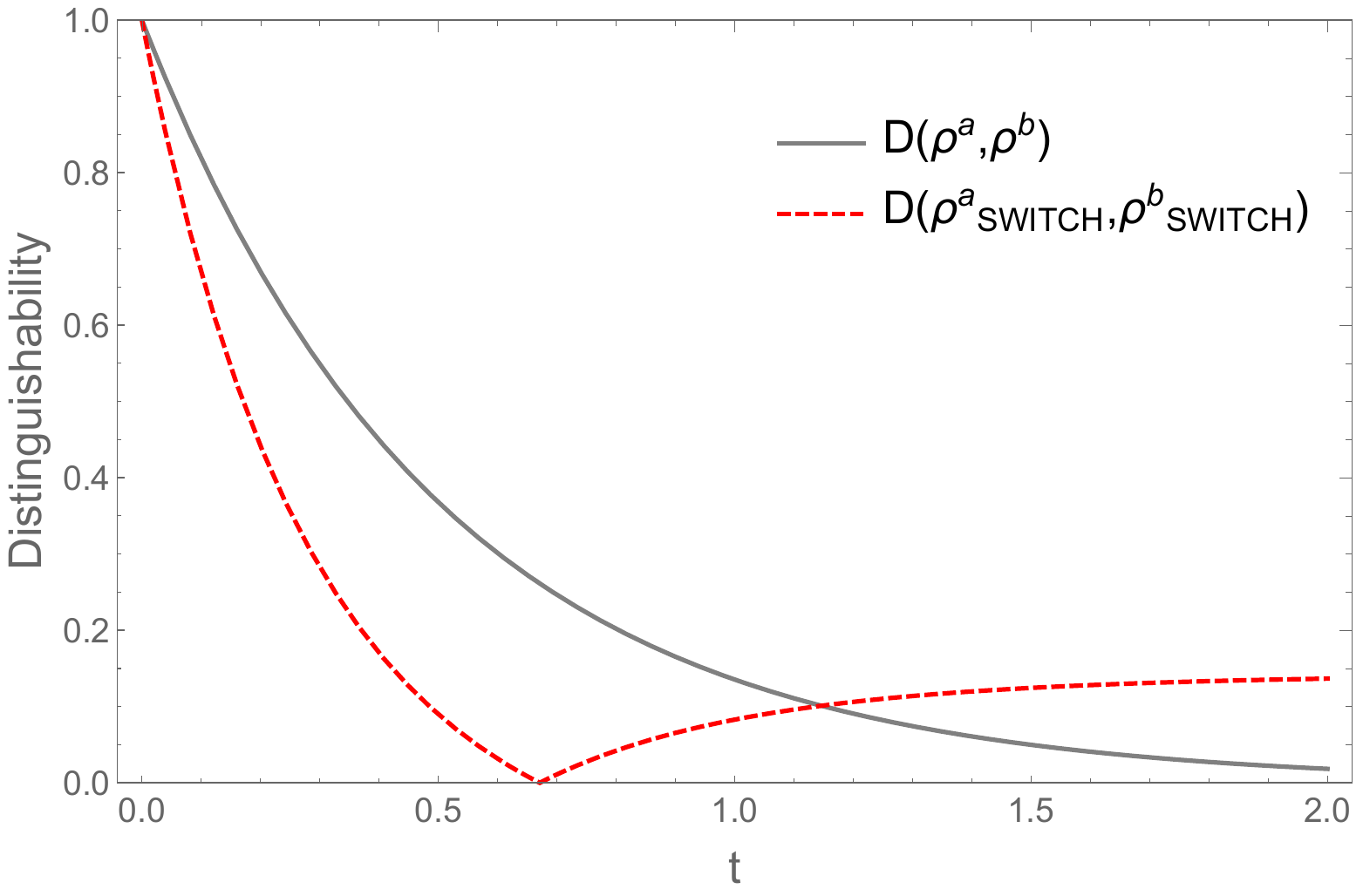}
\caption{The distinguishability $D (\rho^a,\rho^b)$ of two states $\rho^a$ and $\rho^b$ is plotted with time. The solid gray curve represents the monotonically decreasing behaviour of $D (\rho^a,\rho^b)$ for eternal non-Markovian evolution. The dashed red line (obtained after the switching two same channels showing eternal non-Markovianity) shows the non-monotonic behaviour of $D(\rho^a_{\texttt{SWITCH}},\rho^b_{\texttt{SWITCH}})$ representing the activation of eternal non-Markovianity through quantum-SWITCH.}
\label{plot1}
\end{figure}

Let us now take two states $\rho^a = \left(\begin{array}{cc} 1  & 0 \\ 0 & 0 \end{array}\right)$ and $\rho^b = \left(\begin{array}{cc} 0  & 0 \\ 0 & 1 \end{array}\right)$. The distance between two such states can be represented as $D (\rho^a,\rho^b)= \frac{1}{2}||\rho^a -\rho^b ||_1$ where $||X||_1 = \text{Tr}[\sqrt{X^\dagger X}]$ denotes the trace distance. It becomes evident that under this evolution, the distance falls monotonically, reaching zero only after an infinite time (see fig. \eqref{plot1}). Hence, despite the non-Markovian nature of the dynamics, there is an absence of information backflow.

Now the question arises, given exactly similar several copies of such dynamical maps whether we can activate this non-Markovianity such that information back flow can be triggered. In order to do that, the channels may be used in series or in parallel connection. However, one may show that this dynamical map can not be activated even if one uses several copies of such dynamical maps either in series or in parallel when no additional resources are available. For the interested readers, we have explicitly proved this fact in the Appendix \ref{appendix}. Next, one may also ask whether any convex mixture of the action of two copies of such channels in two different causal order can lead to the activation phenomena. To address this question, let's consider a convex combination of causal orders of the action of two general channels say $\mathcal{N}_1$ and $\mathcal{N}_2$ on some target state $\rho$. To illustrate, suppose with probability $p$, the order $\mathcal{N}_2 \circ \mathcal{N}_1$ has been applied on the target state $\rho$ and with probability $(1-p)$, the order $\mathcal{N}_1 \circ \mathcal{N}_2$ has been applied on the state $\rho$. If $\lbrace K_{i}^1 \rbrace$ and $\lbrace K_{j}^2 \rbrace$ represent the set of Kraus operators corresponding to the channel $\mathcal{N}_1$ and $\mathcal{N}_2$ respectively then after the action of quantum SWITCH, the joint state of the target and control system becomes
\begin{widetext}\small
\begin{equation}
    \sum_{i,j} (\sqrt{p}K_j^{(2)} \circ K_i^{(1)} \otimes \ket{0}_c\bra{0} + \sqrt{1-p} K_i^{(1)} \circ K_j^{(2)} \otimes \ket{1}_c\bra{1})(\rho\otimes \omega_c)( \sqrt{p}K_j^{(2)} \circ K_i^{(1)} \otimes \ket{0}_c\bra{0} + \sqrt{1-p} K_i^{(1)} \circ K_j^{(2)} \otimes \ket{1}_c\bra{1})^\dagger
\end{equation}
\end{widetext}
where $\omega_c$ is the initially prepared state for the control qubit. Now in order to have the convex mixture of two causal order, we trace out the control system and have the expression
\begin{align*}
    \sum_{i,j} &p(K_j^{(2)} \circ K_i^{(1)}\rho K_i^{(1)\dagger} \circ K_j^{(2)\dagger} \nonumber \\&+ (1-p)(K_i^{(1)} \circ K_j^{(2)}\rho K_j^{(2)\dagger} \circ K_i^{(1)\dagger}.
\end{align*}
One may note that since $i,j$ are dummy indexes and two channels are exactly same (i.e, $\mathcal{N}_1 = \mathcal{N}_2 = \mathcal{N}$), above expression is nothing but $\sum_{i,j} (K_j^{(2)} \circ K_i^{(1)}\rho K_i^{(1)\dagger} \circ K_j^{(2)\dagger} = \mathcal{N} \circ \mathcal{N} (\rho)$. Therefore any convex mixture of causal order can never activate the information backflow.\\

Let us now exploit these two channels in superposition of two causal orders using an additional control bit. Preparing the initial state of the control bit in $\ket{+}_c\bra{+}$ and after performing the switching action (as presented in the section \ref{s2}), we calculate the evolution of the states $\rho_a$ and $\rho_b$ through the effective channel. The final state can be obtained after measuring the control qubit in the $\lbrace \ket{\pm}_c\bra{\pm}\rbrace$ basis. The general form of the state $\rho$ after the action of quantum SWITCH will look as
\begin{align}\label{switch_map}
& \rho_{\texttt{SWITCH}}  = \frac{1}{n(t)} \left[ {}_c\bra{+} S(\rho \otimes \ket{+}_c\bra{+})\ket{+}_c \right]=\nonumber \\
&\left(\begin{array}{cc} A(t) \rho_{11}(0) + B(t) \rho_{22}(0)  & A(t) \rho_{12}(0) \\ A(t) \rho_{21} (0) & B(t)\rho_{11}(0) + A(t)\rho_{22}(0) \end{array}\right)
\end{align}
with $A(t) = \frac{3+2e^{2t}+3e^{4t}}{-1+2e^{2t}+7e^{4t}}$, $B(t)=\frac{4(-1+e^{4t})}{-1+2e^{2t}+7e^{4t}}$ and $n(t) = \frac{7+2e^{-2t}-e^{-4t}}{8}$ is the success probability of getting $`+'$ outcome.

Now, after calculating the distance between these two states $D(\rho^a_{\texttt{SWITCH}},\rho^b_{\texttt{SWITCH}})$, one can find that the distance does not fall monotonically rather a revival will take place (see fig. \eqref{plot1}). It can thus be concluded that information backflow has occurred after the switching action even if the original channel does not show any information backflow. Since an exactly same channel is used twice in order to activate the information backflow phenomenon and no additional channels are used, we say it to be self activation.

Before delving into the detailed analysis of the switch induced dynamical map, let us first clarify its divisibility aspect. The CP divisibility of a quantum dynamical map as defined earlier, can be determined by the eigen-spectrum of the corresponding Choi state, iff the channel is invertible. Moreover, for a invertible dynamical map, the P-divisibility property can be determined by checking whether all intermediate maps of the dynamics are positive or not. However, if the quantum channel under question, is not invertible, these criteria are no longer valid. Interestingly, the switch induced map (Eq. \eqref{switch_map}) violates the invertibility condition \citep{invert1,invert2} at a particular time $t^*$, where $A(t^*)=B(t^*)=\frac{1}{2}$. We call this time, $t^*$ to be the {\it characteristics time} that has been discussed elaborately in section~\ref{s4}. At this point, the output of the map is maximally mixed state irrespective of any initial state. It is straight forward to check that such maps are not invertible. For such non-invertible dynamical maps the criteria for verifying divisibility of the operation changes \citep{invert1,invert2,wolf,Davalos}.\\

If we have a non-invertible Pauli channel $\Lambda(.)$ and $\Lambda(\sigma_i)=\lambda_i\sigma_i$ where $\sigma_i$s represent Pauli matrices, then the map is CP divisible if there is only one $k\in \{1,2,3\}$, for which $\lambda_k \neq 0$ \citep{wolf,invert2}. For the switched dynamical map in \eqref{switch_map}, we find that $\lambda_1=\lambda_2=B(t)$, $\lambda_3=A(t)-B(t)$. It is straight forward to check that $B(t)\neq 0$ except at $t=0$. Therefore $\lambda_1$ and $\lambda_2$ remains non zero always, proving the map is not CP divisible. Furthermore, the condition for retaining P-divisibility for a non-invertible Pauli channel is $\lambda_1\lambda_2\lambda_3 \geq 0$ \citep{wolf,invert2}. For the switch-induced dynamical map (of \eqref{switch_map}), the condition therefore becomes: $B(t)^2(A(t)-B(t))\geq 0$, or $B(t) \leq A(t)$. It can be directly checked from the following plot that this condition is violated when ever $B(t) \geq A(t)$, proving the map is not also P-divisible.

\begin{figure}[htb]
\centering
\includegraphics[height=5cm,width=8cm]{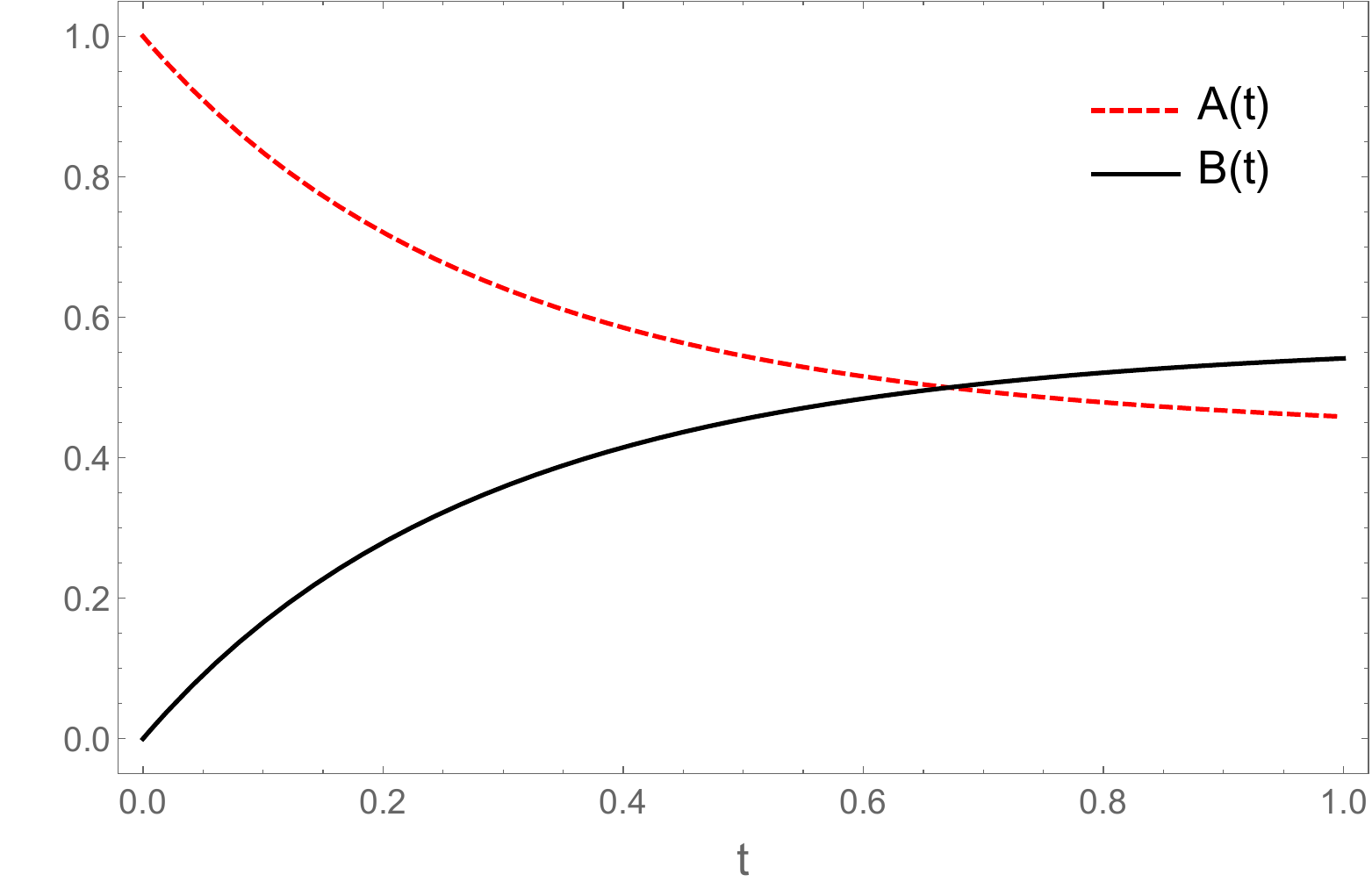}
\caption{$A(t)$ and $B(t)$ are plotted with respect to $t$ to check when the condition for P-divisibility is breaking.}
\label{plot3}
\end{figure}

\section{Dynamics of the activation of Eternal non-Markovianity}\label{s4}
We are now interested to see the dynamics when switching action has been taken place. For this, consider the initial state to be $\rho(0)$ and after the switching action the state becomes $\rho(t)$. The dynamical map for the switch induced evolution is represented by Eq.\eqref{switch_map}, in the previous section.

Let us now derive the exact canonical master equation of the Lindblad type for the dynamical map generated by quantum SWITCH. Here it might be noted that after the action of quantum SWITCH, the control bit is finally measured (say in $\lbrace \ket{+}\bra{+}, \ket{-}\bra{-}\rbrace$ basis) and although there is a certain probability of obtaining each outcome ($`+',`-'$), we are considering dynamics corresponding to a particular outcome only. Therefore, apparently it may seems difficult to derive a Lindblad type master equation since the trace preservation condition is not maintained. However, one may also note that in our case, the probability of obtaining $`+',$ or $`-'$ is independent of the initial state rendering the dynamics linear and trace preserving upto that probabilities and hence the final state \eqref{switch_map} depends linearly on the initial state. Consideration of the post-selected state is also justifiable since in quantum switch mechanisms, the advantages are primarily showcased when focusing on conditional or post-selected states determined by the measurement outcomes. However, the average state obtained after tracing out the control, does not yield any advantage. 

Let us now represent the dynamical map as
\begin{equation}\label{dynamicalmap}
    \rho (t) = \Omega [\rho (0)].
\end{equation}
The equation of motion will be of the form
\begin{equation}\label{eqom}
    \Dot{\rho} (t) = \Tilde{\Lambda} [\rho (t)]
\end{equation}
where $\Tilde{\Lambda} [.]$ is the generator of the dynamics. Now following Ref. \cite{Hall14,Bhattacharya17}, we can find the master equation and generator of the dynamics. 

Here $\lbrace \mathcal{G}_i\rbrace$ denotes the orthonormal basis set with the properties $\mathcal{G}_0 = \mathbb{I}/\sqrt{2}$, $\mathcal{G}_i^{\dagger} = \mathcal{G}_i$, $\mathcal{G}_i$ are traceless except $\mathcal{G}_0$ and $\text{Tr}[\mathcal{G}_i\mathcal{G}_j] = \delta_{ij}$. The map \eqref{dynamicalmap} can be represented as
\begin{equation}
    \Omega [\rho (0)] = \sum_{m,n} \text{Tr}[\mathcal{G}_m \Omega [\mathcal{G}_n]] \text{Tr}[\mathcal{G}_n \rho (0)] \mathcal{G}_m = [F(t)r(0)] \mathcal{G}^T
\end{equation}
where $F_{mn}=\text{Tr}[\mathcal{G}_m \Omega [\mathcal{G}_n]]$ and $r_n(s) = \text{Tr}[\mathcal{G}_n \rho(s)]$. Taking time-derivative of the above equation, we shall get
\begin{equation}
    \dot{\rho}(t) = [\dot{F}(t)r(0)] \mathcal{G}^T.
\end{equation}
Let us now consider a matrix $L$ with elements $L_{mn} = \text{Tr}[\mathcal{G}_m \Tilde{\Lambda} [\mathcal{G}_n]]$. We can therefore represent Eq.\eqref{eqom} as
\begin{equation}\label{Eq13}
    \dot{\rho}(t) = \sum_{m,n} \text{Tr}[\mathcal{G}_m]\Tilde{\Lambda} [\mathcal{G}_n] \text{Tr}[\mathcal{G}_n \rho (t)] \mathcal{G}_m = [L(t)r(t)] \mathcal{G}^T.
\end{equation}
Comparing above two equations we find
\begin{equation}
    \dot{F}(t) = L(t) F(t) \implies L(t)= \dot{F}(t) F(t)^{-1}.
\end{equation}
One may note that $L(t)$ can be obtained if $F(t)^{-1}$ exists and $F(0) = \mathbb{I}$.

Now considering the dynamics \eqref{switch_map} and the orthonormal basis set $\mathcal{G}_l = \lbrace \frac{\mathbb{I}}{\sqrt{2}}, \frac{\sigma_x}{\sqrt{2}},\frac{\sigma_y}{\sqrt{2}},\frac{\sigma_z}{\sqrt{2}} \rbrace$, we evaluate the explicit form of $L(t)$ matrix to be
\begin{equation}
   L(t) = \left(\begin{array}{cccc} 0  & 0 & 0 & 0\\ 0 & \frac{\dot{A}(t)}{A(t)} & 0 & 0\\ 0 & 0 & \frac{\dot{A}(t)}{A(t)} & 0\\ 0 & 0 & 0 & \frac{\dot{A}(t) - \dot{B}(t)}{A(t) - B(t)}\end{array}\right).
\end{equation}
Using the expression of $L(t)$, Eq. \eqref{Eq13} may be written in the simplified form
\begin{align}
    \dot{\rho}(t) &= L_{xx} \text{Tr}[\mathcal{G}_x \rho (t)] \mathcal{G}_x + L_{yy} \text{Tr}[\mathcal{G}_y \rho (t)] \mathcal{G}_y \nonumber \\
    &~~~~+ L_{zz} \text{Tr}[\mathcal{G}_z \rho (t)] \mathcal{G}_z \nonumber
\end{align}
since other $L_{ij}$ terms vanishes. One may also evaluate that 
\begin{align}
    &L_{xx} \text{Tr}[\mathcal{G}_x \rho (t)] \mathcal{G}_x  = \frac{\dot{A}(t)}{A(t)}\left(\begin{array}{cc} 0  & \frac{\rho_{12}(t)+\rho_{21}(t)}{2} \\ \frac{\rho_{12}(t)+\rho_{21}(t)}{2} & 0 \end{array}\right), \nonumber \\
    &L_{yy} \text{Tr}[\mathcal{G}_y \rho (t)] \mathcal{G}_y  = \frac{\dot{A}(t)}{A(t)}\left(\begin{array}{cc} 0  & \frac{\rho_{12}(t)-\rho_{21}(t)}{2} \\ \frac{\rho_{12}(t)-\rho_{21}(t)}{2} & 0 \end{array}\right), ~\text{and}\nonumber \\
    &L_{zz} \text{Tr}[\mathcal{G}_z \rho (t)] \mathcal{G}_z  = \frac{\dot{A}(t)-\dot{B}(t)}{A(t)-B(t)}\left(\begin{array}{cc} \frac{\rho_{11}(t)- \rho_{22}(t)}{2}  & 0 \\ 0 & \frac{\rho_{22}(t)-\rho_{11}(t)}{2}\end{array}\right). \nonumber
\end{align}

The equation of motion can be written in the simplified form
\begin{align}
    &\dot{\rho}_{11} (t) = \frac{L_{zz}}{2} \rho_{11}(t) - \frac{L_{zz}}{2} \rho_{22}(t), \nonumber \\
    &\dot{\rho}_{12} (t) = L_{xx} \rho_{12}(t),
\end{align}
with $L_{ij}$ being the elements of $L$ matrix. Above equation represents the rate of change of quantum state when the state is transferred through two channels (which initially do not exhibit information backflow) subjected to a quantum SWITCH. 

\begin{figure}[htb]
\centering
\includegraphics[height=4.5cm,width=8cm]{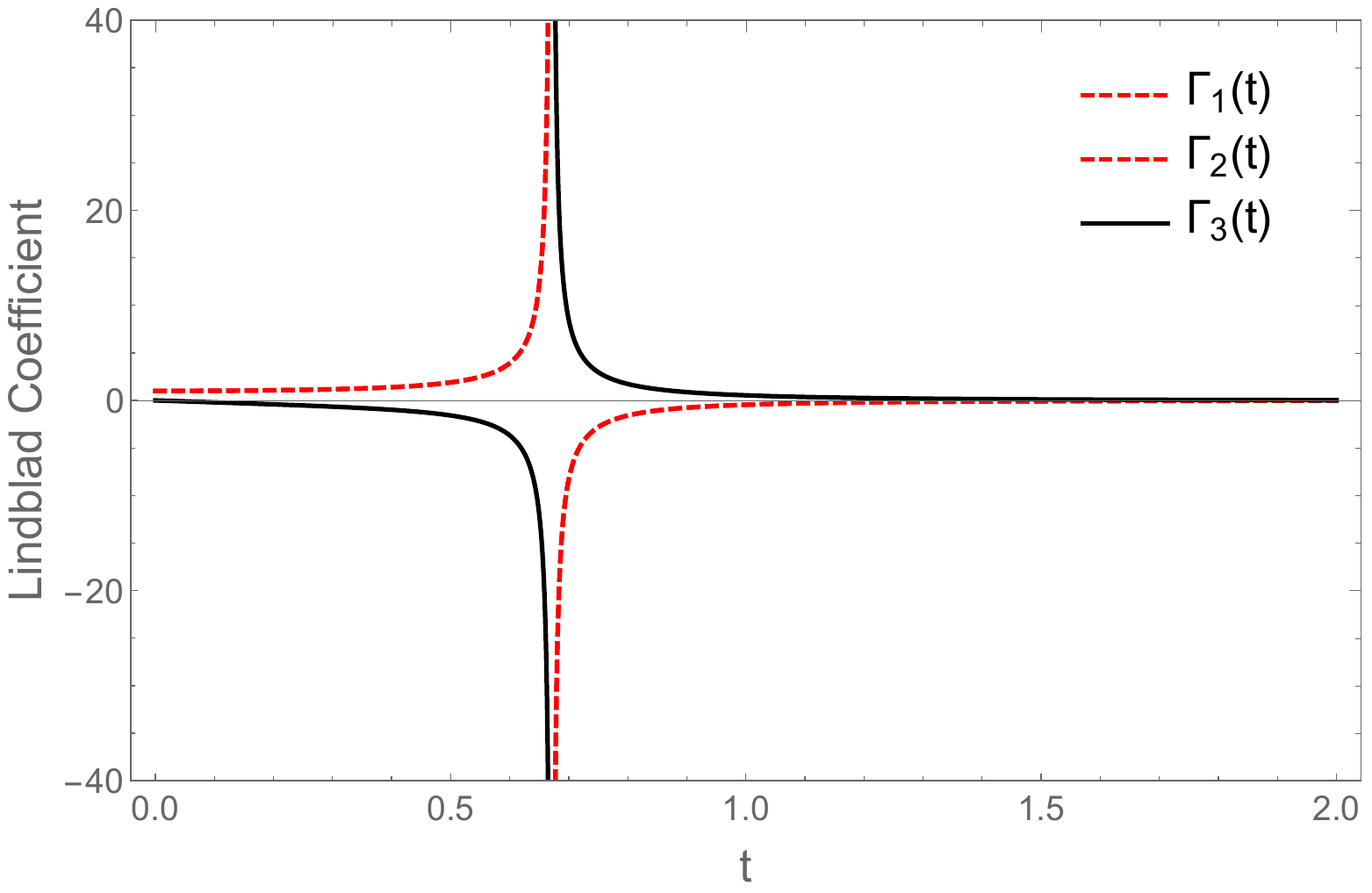}
\caption{The Lindblad coefficient $\Gamma_1 (t), \Gamma_2 (t)$ and $\Gamma_3 (t)$ of the dynamics after the action of quantum SWITCH are plotted with time.}
\label{plot2}
\end{figure}

Let us now derive the Lindblad-type master equation from the above equation. Eq.\eqref{eqom}, can be represented as \cite{Hall14},  
\begin{equation}
    \dot{\rho} (t) = \Tilde{\Lambda} [\rho (t)] = \sum_{l} A_l (t) \rho B_l^{\dagger}(t)
\end{equation}
where $A_l (t) = \sum_m \mathcal{G}_m a_{ml}(t)$ and $B_l (t) = \sum_m \mathcal{G}_m b_{ml}(t)$ with $\lbrace \mathcal{G}_m \rbrace$ are the basis vectors as defined earlier. Now exerting the expression of $A_l (t)$ and $B_l (t)$, above equation can be represented as
\begin{equation}
    \dot{\rho} (t) = \sum_{m,n = 0,x,y,z} z_{mn}(t) \mathcal{G}_m \rho (t)\mathcal{G}_n 
\end{equation}
where $z_{mn}(t) = \sum_l x_{ml}(t)y_{ln}(t)^*$ are the elements of a Hermitian matrix. Let us now represent above equation in the following form
\begin{align}
    \dot{\rho} (t) = &\frac{i}{\hbar} [\rho (t), H(t)] \nonumber \\
    &+ \sum_{m,n = \lbrace x,y,z\rbrace} z_{mn} (t)[\mathcal{G}_m \rho (t) \mathcal{G}_n - \frac{1}{2}\lbrace \mathcal{G}_n\mathcal{G}_m, \rho (t)\rbrace]
\end{align}
where $H (t)= \frac{i}{2} \hbar (\mathcal{H}(t) - \mathcal{H}^{\dagger}(t))$ with $\mathcal{H}(t) = \frac{z_{00}(t)}{8}\mathbb{I} + \sum_m \frac{z_{m0}}{2}\mathcal{G}_i$ and $\lbrace .\rbrace$ represents anticommutator. Therefore, the canonical form of master equation of the Lindblad form will be
\begin{align}\label{n1}
    \dot{\rho} (t) &= \mathcal{L}_t (\rho (t)) \nonumber \\
    &= \Gamma_{1} (t)[\sigma_x \rho (t) \sigma_x - \rho (t)]  + \Gamma_{2} (t)[\sigma_y \rho (t) \sigma_y - \rho (t)] \nonumber \\
    &~~~~+ \Gamma_{3} (t)[\sigma_z \rho (t) \sigma_z - \rho (t)].
\end{align}
This equation represents a general form of the Lindblad type Master equation. However, for the dynamical map generated
by the quantum SWITCH, $\Gamma_1 (t), \Gamma_2 (t)$ and $\Gamma_3 (t)$ are given by

\begin{align}\label{Gamma}
    \Gamma_{\text{1}} (t) &= \Gamma_{\text{2}} (t) = -\frac{L_{zz}}{4}  = -\frac{1}{4}\frac{\dot{A}(t)-\dot{B}(t)}{A(t)-B(t)},\nonumber \\
    \Gamma_{\text{3}}(t) &=   \frac{L_{zz}}{4} - \frac{L_{xx}}{2} =  \frac{1}{4}\frac{\dot{A}(t)-\dot{B}(t)}{A(t)-B(t)} - \frac{1}{2} \frac{\dot{A}(t)}{A(t)}.
\end{align}
The Lindblad coefficients $\Gamma_1 (t), \Gamma_2 (t)$ and $\Gamma_3 (t)$ are plotted with time in fig.\eqref{plot2}. 

The previous discussion has covered the CP-indivisible property of the dynamical map. Let us now focus on the Lindblad-type Master equation for analyzing the activation of information backflow. To investigate this fact in a more detailed manner, we prove the following statements.

\textbf{Statement 1:} The sufficient condition of having no information backflow (i.e, non-increasing trace distance) for two arbitrary qubits undergoing general depolarizing dynamics of the form given in equation \eqref{n1} is given by $\Gamma_i (t)+ \Gamma_j(t)~\geq 0$ for all $i \neq j$.

\proof Let us consider two arbitrary qubits of the form 
\[ \tau_i(t)=\frac{1}{2}\left( \mathbb{I}+\Vec{t}_i(t).\Vec{\sigma}\right)~~\mbox{with}~~i=(1,2), \]
where the time dependent Bloch vectors $\Vec{t}_i=(t_{1}^i(t),t_{2}^i(t),t_{3}^i(t))$ and $\Vec{\sigma}=(\sigma_x,\sigma_y,\sigma_z)$ are represented in the usual form. The trace distance between these two states can be calculated as 
$ D(\tau_1(t),\tau_2(t))=\sqrt{a_1(t)^2+a_2(t)^2+a_3(t)^2}$, with $a_k(t)= (t_k^1(t)-t_k^2(t))$ for $k=(1,2,3)$. The time derivative of the trace distance can thus be calculated as 
\begin{widetext}
\begin{equation} \label{derivative}
    \frac{d}{dt}D(\tau_1(t),\tau_2(t))=   \lim_{\epsilon\to 0}\frac{\sqrt{a_1(t+\epsilon)^2+a_2(t+\epsilon)^2+a_3(t+\epsilon)^2}-\sqrt{a_1(t)^2+a_2(t)^2+a_3(t)^2}}{\epsilon},
\end{equation}  
\end{widetext}
where 
\begin{align} \label{condition_for_backflow}
    a_1(t+\epsilon) &= a_1(t)(1 - 2\epsilon (\Gamma_2(t) + \Gamma_3(t))),  \nonumber \\
    a_2(t+\epsilon) &= a_2(t)(1 - 2\epsilon (\Gamma_1(t) + \Gamma_3(t))), \nonumber \\
    a_3(t+\epsilon)&=a_3(t)\left(1- 2\epsilon(\Gamma_1(t)+\Gamma_2(t))\right).
\end{align}
Usually, when $\Gamma_i(t) + \Gamma_j(t) \geq 0 ~\forall i\neq j$ then the first term of r.h.s of Eq. \eqref{derivative} will always be less than the second term since $\epsilon\Gamma_i(t)<<1$, making $\frac{d}{dt}D(\tau_1(t),\tau_2(t))$ to be negative. Hence, there will be no information backflow as the trace distance is non-increasing. \qed

\textbf{Statement 2:} The necessary condition of information backflow (i.e, increasing trace distance) for general depolarizing dynamics of the form given in equation \eqref{n1} is given by $\Gamma_i (t)+ \Gamma_j(t)~< 0$ for any $i \neq j$.

\proof The derivative of the trace distance (Eq.~\ref{derivative}) and the expression of $a_i(t+\epsilon)$ (Eq.~\ref{condition_for_backflow}) implies that if for any $i\neq j$, $\Gamma_i (t)+ \Gamma_j(t)~< 0$, we can choose initial state such that $a_i=a_j=0$ so that only $\Gamma_i (t)+ \Gamma_j(t)$ term corresponding to negative value can survive. In that case, since other terms are zero and $a_k(t+\epsilon)>a_k(t)$ (where $k\neq i\neq j$), the derivative of the trace distance will be positive indicating information backflow. \qed 

\textbf{statement 1} and \textbf{Statement 2} are quite general to the family of dynamics given by Eq. \eqref{n1}. Below we explicitly analyse \textbf{Statement 2} for the dynamical map generated by quantum SWITCH. For that particular dynamical map, it is straightforward to show (from Eq. \eqref{Gamma}) that
\begin{equation}
    \Gamma_1 (t) +\Gamma_3 (t) = \Gamma_2 (t)+\Gamma_3 (t)= - \frac{\dot{A}(t)}{2A(t)} > 0,
\end{equation}
where the inequality follows from the fact that $A(t) > 0$ while the derivative of $A(t)$ is negative. On the other hand $\Gamma_1(t)+\Gamma_2(t)=-\frac{\dot{A}(t)}{2A(t)-1} \leq 0$ after certain time $t^*$. This obvious fact can be checked from the expression of $A(t)$ and the value of $t^*$ has been calculated latter. Therefore, under the condition $\epsilon\Gamma_i(t)<< 1$, if we choose initial states to be $\rho^a = \ket{0}\bra{0}$ and $\rho^b=\ket{1}\bra{1}$ (which implies only $a_3(t)$ survives), after time $t^*$ the first term of the numerator in the right hand side of the equation \eqref{derivative} will be greater than the second term, since the conditions $\Gamma_1(t) + \Gamma_2(t) < 0$ hold and hence there will be information backflow.

On the otherhand, if we consider the original channel given in \eqref{depol1} with $\gamma_1(t)=\gamma_2(t)=1$ and $\gamma_3(t)=-\tanh{t}$, it can be checked that the positivity of the infinitesimal dynamics is retained with the condition $\tanh{t}\leq 1$, which is always true. It has been shown \citep{Hall14} that for this types of eternal non-Markovian channels, monotones like trace distance and a few others do not show information backflow. This is because although the dynamics is breaking CP-divisibility (and hence non-Markovian), at every instant of time the positivity of the infinitesimal maps are always retained. As we have shown in the previous analysis, by using quantum switch, this particular property can be altered and hence a self activation (activation using same channel) of non-Markovian information backflow can be initiated. 

Fig. \ref{plot1} suggests that after a certain {\it characteristic time} $t^*$, the information backflow prevails. Let us now define and evaluate that {\it characteristic time} for our case. The {\it characteristic time} is the earliest time at which information backflow can be triggered. In order to find an explicit expression for the {\it characteristic time}, one need to look at the cross over time at which the time derivative of $D(\tau_1(t),\tau_2(t))$ changes its sign from $-$ve to $+$ve. Therefore, from Eq. \eqref{derivative}, we need to determine the condition for $\frac{d}{dt}D(\tau_1(t),\tau_2(t))= 0$. After simplification, one can see that above condition holds for
\begin{align}
    \lim_{\epsilon\to 0}\sum_{i=1}^{3} a_i(t^*+\epsilon)^2 -a_i(t^*)^2 = 0.
\end{align}
Since $\epsilon$ is sufficiently small, from the expressions of $a_i(t+\epsilon)$ one may note that above condition can be re-written as $a_1(t^*)^2[-2\epsilon (\Gamma_2(t^*)+\Gamma_3(t^*))] +  a_2(t^*)^2 [-2\epsilon (\Gamma_1(t^*)+\Gamma_3(t^*))]+ a_3(t^*)^2[-2\epsilon (\Gamma_1(t^*)+\Gamma_2(t^*))] = 0$ or $a_1(t^*)^2 [\Gamma_2(t^*)+\Gamma_3(t^*)] + a_2(t^*)^2[\Gamma_1(t^*)+\Gamma_3(t^*)] + a_3(t^*)^2[ \Gamma_1(t^*)+\Gamma_2(t^*)] = 0$.

For our case, we have $\Gamma_1 (t) = \Gamma_2 (t)$ as given in Eq. \eqref{Gamma}. Therefore, above condition implies 
\begin{equation}
    \frac{a_1(t^*)^2 + a_2(t^*)^2}{a_3(t^*)^2} = -\frac{\Gamma_1 (t^*) + \Gamma_2 (t^*)}{\Gamma_1 (t^*) + \Gamma_3 (t^*)}.
\end{equation}
Now since, $a_1 (t) = a_1 (0) A(t)$, $a_2 (t) = a_2 (0) A(t)$ and $a_3 (t) = a_3 (0) (A(t)-B(t))$, above condition suggests
\begin{align}
    \frac{a_1(0)^2 + a_2(0)^2}{a_3(0)^2} \frac{A(t^*)^2}{(A(t^*) -B(t^*))^2} = - \frac{2A(t^*)}{(A(t^*) -B(t^*))}
\end{align}
where $a_1 (0), a_2 (0)$ and $a_3 (0)$ are decided by the initial state. One can further simply to have the condition
\begin{equation}
    2 \frac{A(t^*) -B(t^*)}{A(t^*)} = -\chi (0)
\end{equation}
where, $\chi (0) = \frac{a_1(0)^2 + a_2(0)^2}{a_3(0)^2}$ is a function of initial state parameter. Solution of this equation gives rise to the {\it characteristic time} i.e, the earliest time at which information backflow can be triggered which however depends on the initial state parameter $\chi (0)$.

For our particular case, the initial state has been considered as $\tau_1(0) = \left(\begin{array}{cc} 1  & 0 \\ 0 & 0 \end{array}\right)$ and $\tau_2(0) = \left(\begin{array}{cc} 0  & 0 \\ 0 & 1 \end{array}\right)$. For that case the condition for earliest information backflow reduces to 
\begin{equation}
    A(t^*) -B(t^*) = 0
\end{equation}
Solving above equation, one may evaluate the {\it characteristic time}, $t^* = \frac{1}{2}\ln{(1+2\sqrt{2})} \approx 0.67$ which perfectly matches with Fig. \ref{plot2}.
\section{Discussions}
Certain evolution or dynamical maps which although are non-Markovian, do not attribute information backflow. For such dynamics, the non-Markovian feature is not captured by the usual measure of non-Markovianity based on information backflow. We investigate that information backflow can never be activated for those channels even when many copies of such channels are used in series or in parallel combination. However, exploiting two such channels in superposition of different orders, we find that information backflow can be restored. We then find out actual cause of this phenomenon of activation of information backflow by looking at the reduced dynamics of the effective switched channel. We show that though the original invertible dynamics, is not CP-divisible, but it always preserve P-divisibility and hence is unable to generate information backflow. However, for the switched evolution, along with its invertibility, both CP-divisibility and P-divisibility of the dynamics also breaks down. Moreover, the presence of the switching action also leads to the activation of information backflow. 

Before concluding, a few remarks are in order. In this work, we are neither demanding that the way of activating information backflow presented in this paper is the only procedure to activate such things nor activation of information backflow is impossible by any higher order quantum process \cite{Chiribella08(1)}. Therefore, further study regarding the activation of information backflow using causally ordered quantum comb \cite{Chiribella08(1)} are in order. On the other hand, finding out the fundamental cause behind the activation of quantum channel exploiting indefinite causal order is one of the most important open problems which are yet to be addressed. Although in this paper, we address the question partially, the question remains open for the advantages of indefinite causal order in other tasks especially for quantum communication. Despite the recent attempts to establish the connection between non-Markovianity and causal nonseparability \cite{Milz_2018,UShrikant}, this work explores the dynamical perspectives of indefinite causal order from the backdrop of Lindblad type evolution. Since both indefinite causal order and non-Markovianity are fundamentally related to the memory of quantum systems, further investigations are in order, to decipher the connections between these two novel phenomena.

\section{acknowledgement} 
AGM thankfully acknowledges the academic visit during early May ($2022$) to the International Institute of Information Technology, Hyderabad where the work has been initiated.
\bibliography{Hidden_NM}

\onecolumngrid
\appendix
\section{For invertible dynamical maps, information backflow can not be activated by series and parallel action of the channels}\label{appendix}
The fact that information backflow can not be activated by series and parallel action of the channels can be proved through the following statement.\\
\textbf{Statement 3:} Since, the invertible dynamical map given by Eq.\eqref{depol1} does not show information backflow, the usage of several copies of such dynamical maps either in series or in parallel will never lead to the activation of information backflow when no additional resources are allowed.

\proof Let us consider two dynamical maps $\Lambda (t_1^{'},t_0)$ and $\Lambda (t^{'},t_1^{'})$ for some $t^{'}>t_1^{'}>t_0$. In case of series action of two channels, the second channel will act on the output state of the first channel and the second channel being a valid physical map it should be reset again from some initial time say $t_0$, and hence $\Lambda (t^{'},t_1^{'}) \equiv \Lambda (t^{'}-t_1^{'}-t_0,t_0)$.

If we consider $t_1^{'} = t$ and $t^{'} = 2t$ and set $t_0 =0$, the two maps will look like $\Lambda (t_1^{'},t_0) \equiv \Lambda (t,0)$ and $\Lambda (t^{'},t_1^{'}) \equiv \Lambda (t,0)$. The series connection is different than usual concatenation of two maps $\Lambda (t,t^{'}) \circ \Lambda (t^{'},0) = \Lambda (t,0)$ because in that case, the latter map, $\Lambda (t,t^{'})$ may not be CP. Therefore, for the series connection
\begin{align}
    \Lambda (t,0)(\Lambda (t,0)\rho (0))
    &=e^{\int_0^{t}\mathcal{L}(\rho (t))dt}e^{\int_0^{t}\mathcal{L}(\rho (t))dt} \rho (0) \nonumber \\
    &=e^{2\int_0^{t}\mathcal{L}(\rho (t))dt} \rho (0)= e^{\int_0^{t}\mathcal{L}^{'}(\rho (t))dt} \rho (0)
\end{align}
where $\mathcal{L}^{'} (\rho (t))= 2\mathcal{L}(\rho (t))$. Therefore,
\begin{align}
    \frac{d\rho}{dt} = \mathcal{L}^{'} (\rho (t)) = 2\sum_{i=0}^{3} \frac{\gamma_i(t)}{2}[\sigma_i \rho (t) \sigma_i - \rho (t)].
\end{align}
Now we need to show that above dynamics actually retains P-divisibility. For that consider small time approximation of the dynamics, $\Lambda (t + \epsilon, t) (\rho (t)) = (\mathbb{I}+ \epsilon \mathcal{L}^{'}) (\rho (t))$.
In fact, here we are taking the Taylor series expansion of the total dynamical map $\exp{\left(\int_0^t\mathcal{L}^{'}(\rho(t))dt\right)}$, up to first order. 
Hence checking the positivity of this infinitesimal map will be sufficient to determine that whether all possible intermediate maps are positive or not. Now, for checking positivity of this infinitisimal dynamical map, it is sufficient to check the positivity of the map over all possible pure states. Let us take the general form of a pure state $\ket{\phi} = \left(\begin{array}{c} \phi_1   \\ \phi_2  \end{array}\right)$. Then the action of the map on the state $\ket{\phi}\bra{\phi}$ will look as
\begin{align}\small \label{matrix}
\left(\begin{array}{cc} (1- \epsilon [\Gamma_1(t) + \Gamma_2 (t)])|\phi_1|^2 + \epsilon [\Gamma_1 (t)+ \Gamma_2(t)]|\phi_2|^2  & (1- \epsilon [\Gamma_1(t) + \Gamma_2 (t)+ 2 \Gamma_3(t)])\phi_1\phi_2^* + \epsilon [\Gamma_1(t) -\Gamma_2(t)]\phi_1^*\phi_2 \\ (1- \epsilon [\Gamma_1(t) + \Gamma_2(t) + 2 \Gamma_3(t)])\phi_1^*\phi_2 + \epsilon [\Gamma_1(t) -\Gamma_2(t)]\phi_1\phi_2^* & (1- \epsilon [\Gamma_1 (t)+ \Gamma_2(t)])|\phi_2|^2 + \epsilon [\Gamma_1(t) + \Gamma_2(t)]|\phi_1|^2 \end{array}\right)   
\end{align}
The P-divisibility of the map is determined by the positivity of the output matrix presented in the previous equation, for any arbitrary state $\ket{\phi}$. The positivity of an arbitary matrix can be determined by the Sylvestor's criteria \citep{pos1}, which states that any matrix is positive semi-definite, if all of its principal minors are non-negative. For our case of a $2\times 2$ matrix, the principal minors are its diagonal elements and the determinant. Now one can check that diagonal elements are always positive since $\epsilon$ is very small. In order to check whether the map is positive or not, it is now sufficient to check the positivity of the determinant of the above matrix for any value of $\phi_1$ and $\phi_2$. Below we explore under which condition determinant of the above matrix will not be positive. The determinant of the matrix can be calculated as
\begin{align} \label{det}
  &  \epsilon [\Gamma_1 (t)+ \Gamma_2(t)]\lbrace 1-\epsilon [\Gamma_1(t) + \Gamma_2(t)]\rbrace(|\phi_1|^2 - |\phi_2|^2)^2 \nonumber 
     + |\phi_1|^2|\phi_2|^2\lbrace 1- (1-2\epsilon[\Gamma_2(t)+\Gamma_3(t)])^2\rbrace\\
    & + 2\epsilon[\Gamma_1(t)-\Gamma_2(t)](1-\epsilon[\Gamma_1(t)+\Gamma_2(t)+\Gamma_3(t)])\left[|\phi_1|^2|\phi_2|^2-Re(\phi_1\phi_2^*)^2\right],
\end{align}
where ``$Re$" stands for real part of the quantity. It might be noted here that since for our case $\Gamma_1 (t) = \Gamma_2(t)$, the last term of the determinant will not survive. Considering the complete positivity of the overall dynamics and under the condition $\epsilon\Gamma_i (t)<<1~~(\forall i)$, the above term will be positive under the conditions
\[(\Gamma_i (t)+ \Gamma_j(t))~\geq 0~~\forall i \neq j.\]

Therefore, above conditions will always be retained leading to no information backflow (eventually all the Lindblad coefficients are just scaled by a factor of $2$ from the original dynamics).

For the case of parallel connection, two or more channels will act parallelly on different copies of the initial state. The parallel action of the channel can be represented as $(\Lambda (t,t_0) \otimes \Lambda (t,t_0))(\rho_1 (0) \otimes \rho_2 (0))$. Here we only allow separable states on which the map is supposed to act. The reason behind excluding entangled states is that entanglement in itself as a nonlocal correlation can act as a resource to create some information backflow, whereas for separable states, this possibility is excluded. Therefore, under this restriction we can state the following. It is evident $\Lambda (t,t_0) \otimes \Lambda (t,t_0)(\rho_1(0) \otimes\rho_2(0))=\Lambda (t,t_0))(\rho_1 (t_0))\otimes\Lambda (t,t_0))(\rho_2 (t_0))$ will not show any information backflow, because it is acting on a product state. Now from linearity of the operation, this statement can be generalised to the separable states. Hence we can say $\sum_i p_i (\Lambda (t,t_0) (\rho_1^i (t_0)) \otimes \Lambda (t,t_0))(\rho_2^i (t_0))$ will also not show any information backflow. Therefore, when the P-divisible map acts on both sides of a separable state, no information backflow can be observed. Here, only exception might happen, if the maps are applied parallely on an entangled state. This can be seen from the Theorem 1 of Benatti {\it et.al} \citep{benetti}, which states that any one parameter family of maps $\{\Phi_t\}$ is CP divisible iff $\{\Phi_t\otimes\Phi_t\}$ is P-divisible. Since our map $\Lambda (t,t_0)$ is not CP-divisible, $\Lambda (t,t_0)\otimes\Lambda (t,t_0)$ is not P-divisible and hence when acted upon entangled states, information backflow can be observed. However, by {\bf Statement 3}, no entangled states are allowed. Hence, acted upon only on separable states, the map $\Lambda (t,t_0)\otimes\Lambda (t,t_0)$ will not show any information backflow. This proves our claim that parallel action of the  p-divisible channels will also will not show information backflow. A point to remember is that the proof we present here is valid for invertible dynamical maps only. Since we start with such a map, the proof is sufficient for our purpose. The question of non-invertible initial maps, remains open.
\end{document}